\documentclass[12pt]{iopart}
\usepackage{mathptmx}

\usepackage[T1]{fontenc}
\usepackage[latin9]{inputenc}
\usepackage{geometry}
\usepackage{amssymb}
\usepackage{graphicx}

\makeatletter
\@ifundefined{textcolor}{}
{%
 \definecolor{BLACK}{gray}{0}
 \definecolor{WHITE}{gray}{1}
 \definecolor{RED}{rgb}{1,0,0}
 \definecolor{GREEN}{rgb}{0,1,0}
 \definecolor{BLUE}{rgb}{0,0,1}
 \definecolor{CYAN}{cmyk}{1,0,0,0}
 \definecolor{MAGENTA}{cmyk}{0,1,0,0}
 \definecolor{YELLOW}{cmyk}{0,0,1,0}
 }

\@ifundefined{definecolor}
 {\usepackage{color}}{}

\usepackage{mathptmx}

\makeatother

\begin{document}

\title[Coherent control of phase diffusion by scattering
length modulation]{Coherent control of phase diffusion in a Bosonic Josephson junction by scattering
length modulation}

\author{J. Lozada-Vera$^{1,2}$, V. S. Bagnato$^{3}$ and M. C. de Oliveira$^{1,2}$}

\address{$^{1}$Instituto de Física Gleb Wataghin, Universidade Estadual de Campinas,
13083-970 Campinas, SP - Brazil}

\address{$^{2}$Institute for Quantum Information Science, University of Calgary,
Alberta T2N 1N4, Canada}

\address{$^{3}$Instituto de Física de São Carlos, Universidade de São Paulo, 13566-590
São Carlos, SP - Brazil}
\eads{\mailto{{jlozada@ifi.unicamp.br}, \mailto{marcos@ifi.unicamp.br}} }
\begin{abstract}
By means of a temporal-periodic modulation of the $s$-wave scattering
length, a procedure to control the evolution of an initial atomic
coherent state associated with a Bosonic Josephson junction is presented.
The scheme developed has a remarkable advantage of avoiding the quantum
collapse of the state due to phase and number diffusion. This kind
of control could prove useful for atom interferometry using BECs,
where the interactions limit the evolution time stage within the interferometer,
and where the modulation can be induced via magnetic Feshbach resonances
as recently experimentally demonstrated. 
\end{abstract}
\pacs{03.75.Lm,  03.75.Dg, 07.60.Ly}
\maketitle

\section{{\normalsize Introduction}}

A fundamental characteristic of Bose-Einstein condensates (BECs) is
that they display coherence phenomena in analogy to classical waves,
as was observed back in the early experiments, using Young's double
slit \cite{BHE00} and double-well interference settings \cite{A+97}.
The reason being that the order parameter, the macroscopic wave function,
is a complex field with certain amplitude and global phase. Since
then, the importance of measuring the relative phase between fragments
of condensates with high precision has been recognized, and several
different schemes for atom interferometry based in BECs have been
devised \footnote{See for example the recent review in Ref. \cite{G12} and references therein.}. One important result along this line is the creation
of non-classical (entangled) many-body states \cite{SDC+01,EGW+08},
as is the case of the so called squeezed spin states \cite{SS}, exploiting
in a controlled way the nonlinear nature of the atom interactions.
Noteworthy is its application into a new generation of quantum metrology
devices performing below the standard quantum limit \cite{GZN+10}.
However, at the same time, interactions have detrimental effects in
the stage of phase accumulation of the interferometer \cite{GHM+10},
due to a process generally known as {}``phase diffusion'' \cite{phase_diffusion}
blurring the final phase readout and consequently reducing the sensitivity
of the device. A possible way out of this problem might be given by
the management of Feshbach resonances, given the high degree of control
achieved for the different experimental parameters involved with BECs
physics in optical lattices \cite{MO06}. In particular the interatomic
interaction, characterized principally by the \textit{s}-wave scattering
length $a_{s}$, can have both its sign and strength tuned by means
of external magnetic fields, and this has been used extensively in
the proper attainment of BECs \cite{CGJ+10}; in the study of nonlinear
excitations of the condensate, for example in the creation of bright
solitons, for the setting of attractive interactions \cite{bright_solitons};
or in the preparation of an almost ideal Bose gas for Anderson localization
observation \cite{AL}, to name but a few. Another interesting possibility
that has been considered is the modulation in time of the scattering
length. This has proven useful for the control of matter waves, as
in the stabilization of bright and dark solitons \cite{SU03,AKK+03,KTF+03,LZL05},
self-confinement of 2D and 3D BECs without an external trap \cite{ACK+03},
and also for the prediction of the remarkable Faraday pattern formation
in BECs \cite{SLdV02,NC-GK07}. Recently a new set of experiments
have explored this kind of modulation for the generation of turbulence
in BECs \cite{HSR+09}: as it was observed, the scattering length
modulation suppresses the aspect ratio inversion typically observed
in the atomic cloud during free expansion - a signature of the turbulent
regime where the cloud expands freely with constant aspect ratio.

In this work we further investigate the effects of the scattering
length modulation on the quantum dynamics of a BEC trapped in a double
well potential. A typical feature of this system is the existence
of two distinct phases: Josephson oscillations, where atoms tunnel
coherently from one well to the other; and macroscopic self-trapping,
where the interaction between the atoms lead to a halt of the coherent
tunneling mechanism \cite{Albiez+05}. It is well known that in a
full quantum description, an initial atomic coherent state will show
a series of collapse-and-revivals of relative phase and number due
to the atomic interactions and the quantized nature of the matter
wave field \cite{GMH+02}. This is an intrinsic signature of the nonlinear
quantum dynamics, which does not appear in semiclassical mean field
approaches, such as the evolution through a Gross-Pitaevskii equation.
Remarkably, by using the scattering length modulation at certain frequencies,
it is possible to control and avoid the loss of quantum coherence
that takes place during the collapsing process. We describe how this
suppression of collapse occurs and establish a possible connection
which might be important for atom interferometry with BECs. In order
to better understand the dynamics of the collapse, we derived a Fokker-Planck
equation for the Husimi distribution whose evolution can be visualized
on the Bloch sphere parameterizing the relative phase and population
difference variables into the usual angular variables of a spherical
coordinate system. The paper is divided as follows. In Sec. II we
present the quantum mechanical model for the two-mode condensate with
scattering length modulation and show some typical and relevant states.
In Sec. III we discuss the typical collapse and revival of phase and
population dynamics present for a static scattering length. In Sec.
IV we present the central results concerning the use of dynamical
scattering length in the control of coherence of BECs, and finally
in Sec. V we present our conclusions.

\section{{\normalsize Model}}

In order to describe in a full quantum mechanical way an interacting
Bose gas trapped in a double-well potential $V\left(\mathbf{r}\right)$,
at zero temperature, we use the two-mode Bose-Hubbard model \cite{MCW+97,DG11}
characterized by the Hamiltonian ($\hbar\equiv1$):

\begin{eqnarray}
\hat{H} & = & -\frac{\Omega}{2}(\hat{a}_{1}^{\dagger}\hat{a}_{2}+\hat{a}_{1}\hat{a}_{2}^{\dagger})+\epsilon\left(\hat{n}_{1}-\hat{n}_{2}\right) +\kappa\left(t\right)\left[\hat{n}_{1}\left(\hat{n}_{1}-1\right)+\hat{n}_{2}\left(\hat{n}_{2}-1\right)\right],\label{eq:2B-H_Hamiltonian}
\end{eqnarray}
 where $\hat{a}_{j}$ and $\hat{a}_{j}^{\dagger}$ are the annihilation
and creation operators for particles in the site $j=1,2$, with an
associated spatial wave function $\chi_{j}(\mathbf{r})=\langle\mathbf{r}|\chi_{j}\rangle$
with $|\chi_{j}\rangle=\langle\mathbf{r}|\hat{a}_{j}^{\dagger}|\textrm{vac}\rangle$,
satisfying the bosonic algebra $[\hat{a}_{j},\hat{a}_{k}^{\dagger}]=\delta_{jk}$,
and $\hat{n}_{j}=\hat{a}_{j}^{\dagger}\hat{a}_{j}$ the corresponding
number operators. The tunneling frequency or coupling between sites
is defined by $\Omega$, the on-site interaction energy between two
atoms in a single well is given by $\kappa$, which is proportional
to the atomic \textit{s}-wave scattering length $a_{s}$ (with the
usual positive/negative sign convention for repulsive/attractive interaction),
and finally $\epsilon$ is a possible energy offset between the wells
caused by an additional external potential, which in this work will
be considered equal to zero, i.e. a symmetric trap.

As stated in the introduction, we are interested in the case where
$a_{s}$ is periodically modulated via a magnetic Feshbach resonance,
so it is composed of an static as well as a dynamic part. To see how
this can be done, a relation for $a_{s}$ as a function of the magnetic
field $B$ obtained by Moerdijk\textit{ et al}. \cite{MVA97} can
be used:

\begin{equation}
a_{s}\left(B\right)=a_{\textrm{BG}}\left(1-\frac{\Delta}{B-B_{\infty}}\right),
\end{equation}
 with $a_{\textrm{BG}}$ the background or off-resonant value of the
scattering length, $B_{\infty}$ the position of the resonance where
$a_{s}\rightarrow\pm\infty$, and $\Delta$ the resonance width, determined
by the condition $B=B_{\infty}+\Delta$ at which there are no interactions
($a_{s}=0$). In the case of a harmonic magnetic field $B\left(t\right)=\bar{B}+\delta B\cos\omega t$,
and far from the resonance so $\delta B\ll\left|B_{\infty}-\bar{B}\right|$
\footnote{This is relevant since around the resonance region
the loss of atoms is strongly enhanced due to inelastic collisions,
see for example the case of a Na BEC in Ref. \cite{IAS+98}.}, it is obtained at first order in $\delta B$, 
\begin{equation}
a\left(t\right)\simeq\bar{a}+\delta a\cos\omega t,
\end{equation}
 where $\bar{a}=a_{s}(\bar{B})$ and $\delta a=a_{BG}\Delta\delta B/\left(B_{\infty}-\bar{B}\right)^{2}$.
With this expression for the scattering length, the on-site interaction
parameter will have the form $\kappa(t)=\kappa_{0}(1+\mu\cos\omega t)$,
with $\kappa_{0}$ being the static part of the interaction, and $\mu\equiv\bar{a}/\delta a$,
the relative amplitude of the modulation.

To give an idea of actual experimental values, consider for instance
the work of Pollack \textit{et al}. \cite{PDH+10}, where it was used
an ultracold gas of $^{7}$Li atoms, which display a Feshbach resonance
at $B_{\infty}\simeq737$ G. For $\bar{B}=565$ G and $\delta B=14$
G, they obtained an $\bar{a}\simeq3.0a_{0}$ with an amplitude of
modulation $\delta a\simeq2.3a_{0}$. Other possibilities could be
$^{85}$Rb (with a resonance at $B_{\infty}\simeq155$ G), for which
the scattering length is $\bar{a}\sim33a_{0}$ around $\bar{B}\sim165$
G before becoming negative for larger values of $B$, in this case
for example, it could be used values of $\delta B\sim0.2$ G to produce
a modulation of about 30\%; or $^{39}$K (with a resonance at $B_{\infty}\simeq402$
G), for which $\bar{a}\sim6a_{0}$ at $\bar{B}\sim360$ G and negative
for $B<350$ G, and the same modulation can be achieved with $\delta B\sim2$
G \cite{CGJ+10}.

A natural basis for this system is given by the Fock states, 
\begin{equation}
\left|n\right\rangle _{1}\otimes\left|N-n\right\rangle _{2}\equiv\left|n\right\rangle =\frac{(\hat{a}_{1}^{\dagger})^{n}(\hat{a}_{2}^{\dagger})^{N-n}}{\sqrt{n!\left(N-n\right)!}}\left|\textrm{vac}\right\rangle ,\label{eq:fock_states}
\end{equation}
 labeled by $n=0,1,...N,$ the particle number in one of the wells.
They are fragmented states in the sense that the bosons occupy two
different spatial modes (with the particular exception of the states
$\left|0\right\rangle $ and $\left|N\right\rangle $ in which all
bosons are in one mode only). This basis expands an $\left(N+1\right)$-dimensional
Hilbert space which is easily accessible to numerical calculations.
Evidently the total number of particles is conserved as can be seen
directly from the Hamiltonian (\ref{eq:2B-H_Hamiltonian}), i.e. $\hat{N}=\hat{n}_{1}+\hat{n}_{2}$
is a constant of motion. A general $N$-boson state ket is then written
simply as 
\begin{equation}
\left|\psi\left(t\right)\right\rangle =\sum_{n=0}^{N}c_{n}\left(t\right)\left|n\right\rangle ,\label{eq:general_state_fock}
\end{equation}
 with $\sum_{n}\left|c_{n}\left(t\right)\right|^{2}=1$, the normalization
condition for the time-dependent amplitudes. The $\left|n\right\rangle $
states are also commonly known as spin states, since this model is
isomorphic with the SU(2) group through the Schwinger pseudospin operators
\cite{S65} 
\begin{eqnarray}
\hat{J}_{x} & = & \frac{1}{2}(\hat{a}_{1}^{\dagger}\hat{a}_{2}+\hat{a}_{1}\hat{a}_{2}^{\dagger}),\nonumber \\
\hat{J}_{y} & = & \frac{1}{2i}(\hat{a}_{1}^{\dagger}\hat{a}_{2}-\hat{a}_{1}\hat{a}_{2}^{\dagger}),\\
\hat{J}_{z} & = & \frac{1}{2}(\hat{n}_{1}-\hat{n}_{2}),\nonumber 
\end{eqnarray}
 with total angular momentum $J=N/2$ and satisfying the SU(2) angular
momentum algebra $[\hat{J}_{i},\hat{J}_{j}]=i\epsilon_{ijk}\hat{J}_{k}$.
Clearly the $\left|n\right\rangle $ states are eigenstates of the
$\hat{J}^{2}$ and $\hat{J}_{z}$ operators, and then our system can
be seen in terms of a giant spin system with a dynamics governed by
the Hamiltonian 
\begin{equation}
\hat{H}=-\Omega\hat{J}_{x}+2\epsilon\hat{J}_{z}+2\kappa\left(t\right)\hat{J}_{z}^{2}.\label{eq:HamJ}
\end{equation}
 Note that $\hat{J}_{z}=\hat{n}-N/2$, is exactly the relative number
operator, its mean value giving the population imbalance between the
wells.

From the Fock states, a useful basis for the analysis of our problem
can be defined, given by the called \textit{binomial} or \textit{atomic
coherent} \textit{states}, introduced by Arecchi \textit{et al}. \cite{ACG+72}.
They are the more general unfragmented states, which means that all
the bosons are in the same single-particle mode, $\cos(\theta/2)\left|\chi_{1}\right\rangle +e^{i\varphi}\sin(\theta/2)\left|\chi_{2}\right\rangle $,
where $\theta$ and $\varphi$ determine the amplitude and relative
phase, respectively, of single modes $\left|\chi_{i}\right\rangle $,
$(i=1,2)$. Defining $\hat{b}_{1}^{\dagger}=\cos(\theta/2)\hat{a}_{1}^{\dagger}+e^{i\varphi}\sin(\theta/2)\hat{a}_{2}^{\dagger}$
as the creation operator of a particle in the single particle state
the binomial states are defined as 
\begin{equation}
\left|\theta,\varphi\right\rangle =\frac{(\hat{b}_{1}^{\dagger})^{N}}{\sqrt{N!}}\left|\textrm{vac}\right\rangle ,\label{eq:bin_states}
\end{equation}
 or equivalently as 
\begin{equation}
\left|\theta,\varphi\right\rangle =\sum_{n=0}^{N}\sqrt{{N}\choose{n}}\cos^{n}(\theta/2)\sin^{N-n}(\theta/2)e^{i(N-n)\varphi}\left|n\right\rangle ,\label{eq:coherent_state}
\end{equation}
 where the binomial expansion was used. It is worth notice the two
special cases: $\left|0,\varphi\right\rangle $ and $\left|\pi,\varphi\right\rangle $
($\varphi$ undefined), corresponding to the $\left|N\right\rangle $
and $\left|0\right\rangle $ states respectively, which are the only
Fock-coherent states in this description.

\begin{figure}
\begin{centering}
(a)\qquad{}\qquad{}\qquad{}\qquad{}(b)\qquad{}\qquad{}\qquad{}\qquad{}(c) 
\par\end{centering}

\begin{centering}
\medskip{}

\par\end{centering}

\begin{centering}
\includegraphics[scale=0.32]{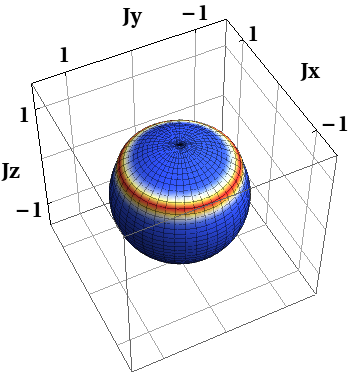}\includegraphics[scale=0.32]{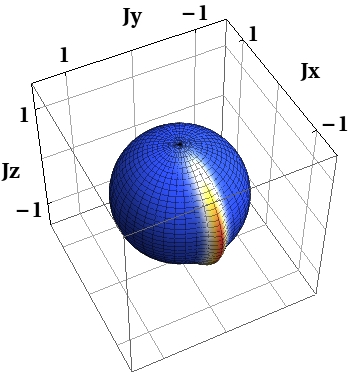}\includegraphics[scale=0.32]{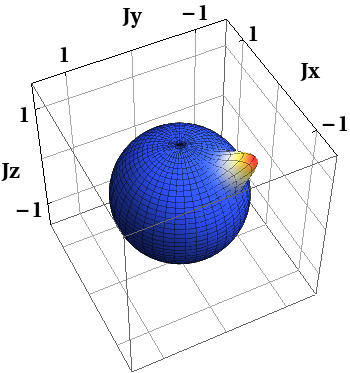} 
\par\end{centering}

\caption{\label{fig:Hexamples}(color online) Examples of Husimi distributions
for three different state preparations with $N=100$: (a) Fock state
$\left|80\right\rangle $; (b) Phase state $\left|\pi\right\rangle $;
(c) Coherent state $\left|\pi/4,5\pi/4\right\rangle $. The $\theta$
and $\varphi$ angles correspond to the usual convention in polar
spherical coordinates.}
\end{figure}

Another basis that can be constructed from the Fock states (\ref{eq:fock_states}),
is given by the Pegg and Barnett \textit{relative phase states} \cite{PB89}.
This is a useful basis since links the results of BEC interferometry
experiments with the two-mode Bose-Hubbard model, providing a way
of describing the phase distribution of a particular state. They are
defined in terms of the Fock states as 
\begin{equation}
\left|\phi_{m}\right\rangle =\frac{1}{\sqrt{N+1}}\sum_{n=0}^{N}\exp(in\phi_{m})\left|n\right\rangle ,
\end{equation}
 with $\phi_{m}=\phi_{0}+2\pi m/(N+1)$, $m=0,1,\ldots,N,$ a quase-continuum
angular variable, and $\phi_{0}=-\pi$, so $\phi_{m}\in[-\pi,\pi)$.
From this definition is clear that the probability amplitudes $C{}_{\phi}$
of a general state written in this basis: $\left|\psi\left(t\right)\right\rangle =\sum_{n=0}^{N}C{}_{\phi}\left(t\right)\left|\phi_{m}\right\rangle $,
are related to the number amplitudes $c_{n}$ via a discrete Fourier
transform. (For a thorough discussion about the two-mode formalism
see the recent review by Dalton and Ghanbari \cite{DG11}).

The binomial states (\ref{eq:coherent_state} ) allow the use of the
semiclassical $Q$-\textit{Husimi distribution}, which is very convenient
for visualizing the dynamics of the many-body state on a spherical
phase space representation (Bloch sphere), and is given by the quasi-probability
distribution of a general state $\left|\psi\right\rangle $ to be
in a coherent state (\ref{eq:coherent_state}): 
\begin{equation}
Q(\theta,\varphi)=\left|\left\langle \theta,\varphi\left|\psi\right.\right\rangle \right|^{2}.
\end{equation}
 This representation is helpful to interpret the results of a certain
dynamics since we can directly extract information about the relative
phase and relative number distribution of a particular state. For
instance, a Fock state, Fig. \ref{fig:Hexamples}(a), which has a
very definite distribution in relative number has a localized $Q$-distribution
along the $z$ axis (which is related to the mean of the $\hat{J}_{z}$
operator); on the other hand, a relative phase state, Fig. \ref{fig:Hexamples}(b),
has a localized distribution around a particular $\phi$ value. Finally,
a coherent state, Fig. \ref{fig:Hexamples}(c), which is the more
classical quantum state possible is represented by an equal noise
distribution around the central value given by the $(\theta,\phi)$
coordinates in the Bloch sphere.

The numerical results presented in the following two sections were
obtained by solving the Schrödinger equation for the Hamiltonian (\ref{eq:2B-H_Hamiltonian})
written in the Fock basis (\ref{eq:general_state_fock}), yielding
a set of $N+1$-coupled differential equations for the expansion amplitudes
$c_{n}(t)$. Relative phase information is extracted from the number
distribution by taking its discrete Fourier transform. 


\section{Static scattering length: collapse of the population oscillations}

\begin{figure}
\begin{centering}
\qquad{}(a)

\medskip{}
 \includegraphics[scale=0.45]{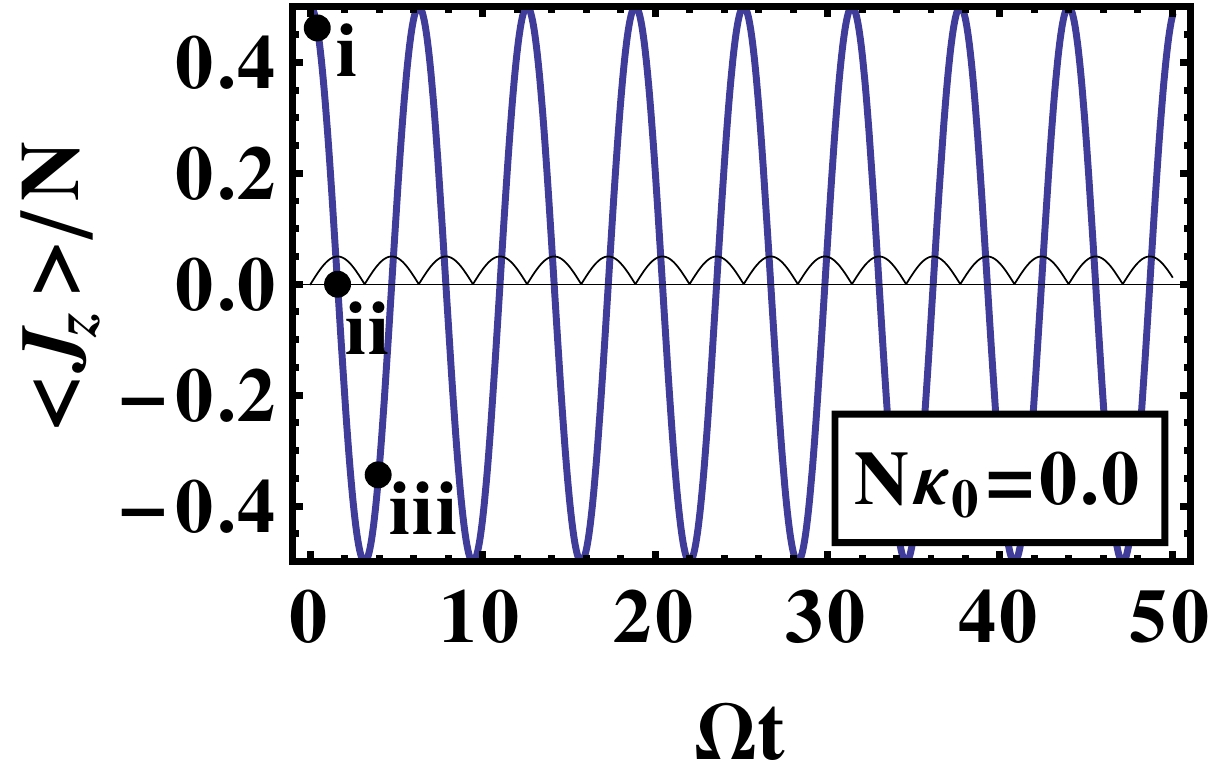}\includegraphics[scale=0.45]{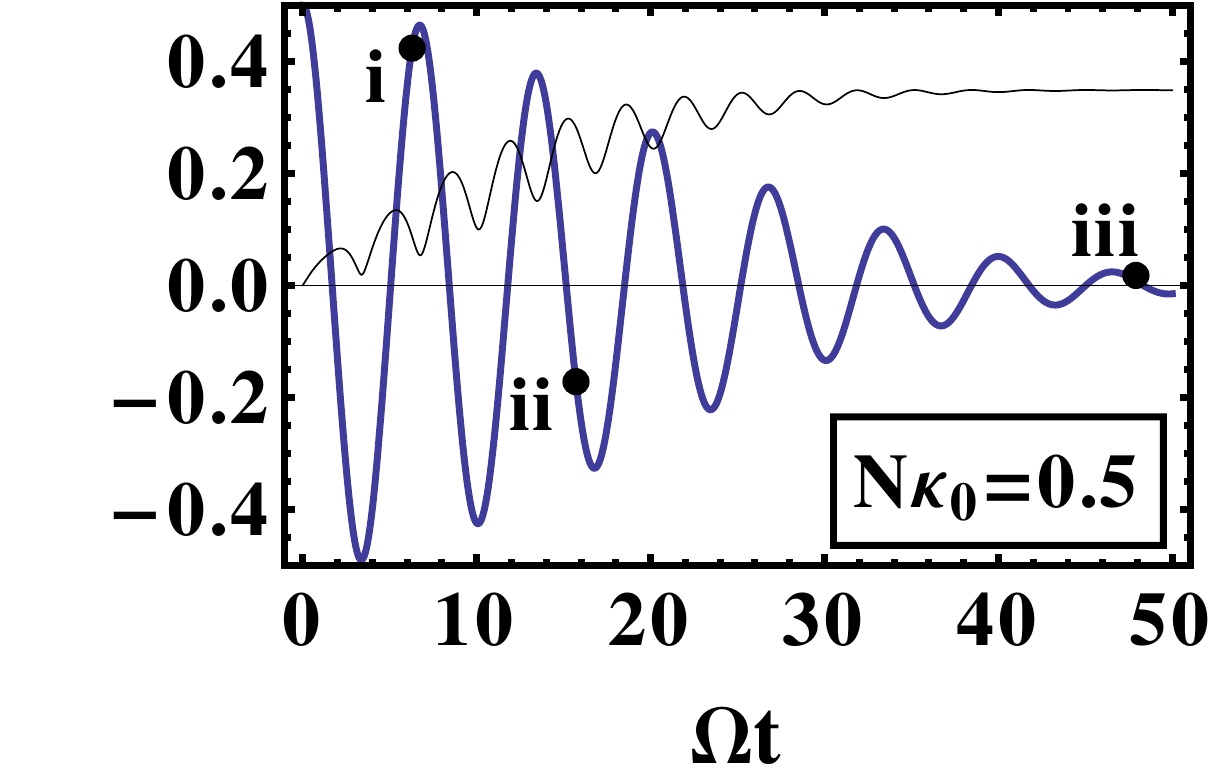}\\
 \medskip{}
 \qquad{}(b) 
\par\end{centering}

\begin{centering}
\qquad{}\enskip{}$\left|c_{n}\right|^{2}$ vs $n$\enskip{}\quad{}\qquad{}\qquad{}\qquad{}\enskip{}\enskip{}\enskip{}$\left|c_{n}\right|^{2}$
vs $n$ 
\par\end{centering}

\begin{centering}
\quad{}\enskip{}\includegraphics[scale=0.4]{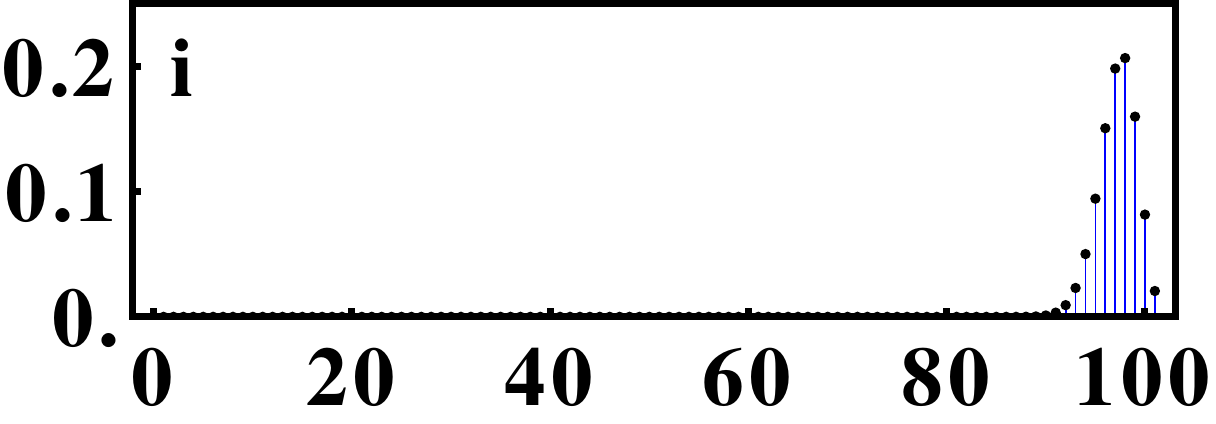}\enskip{}\,\quad{}\includegraphics[scale=0.4]{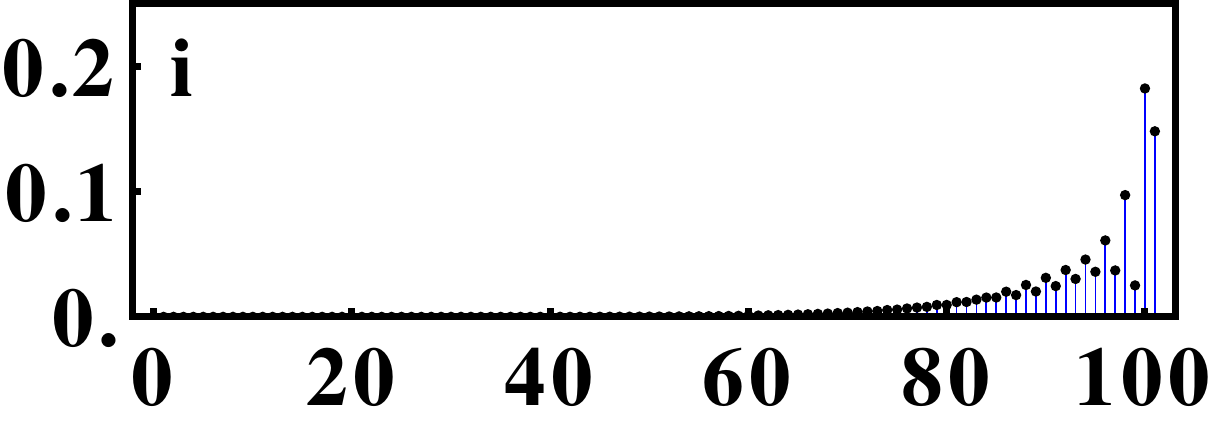} 
\par\end{centering}

\begin{centering}
\quad{}\enskip{}\includegraphics[scale=0.4]{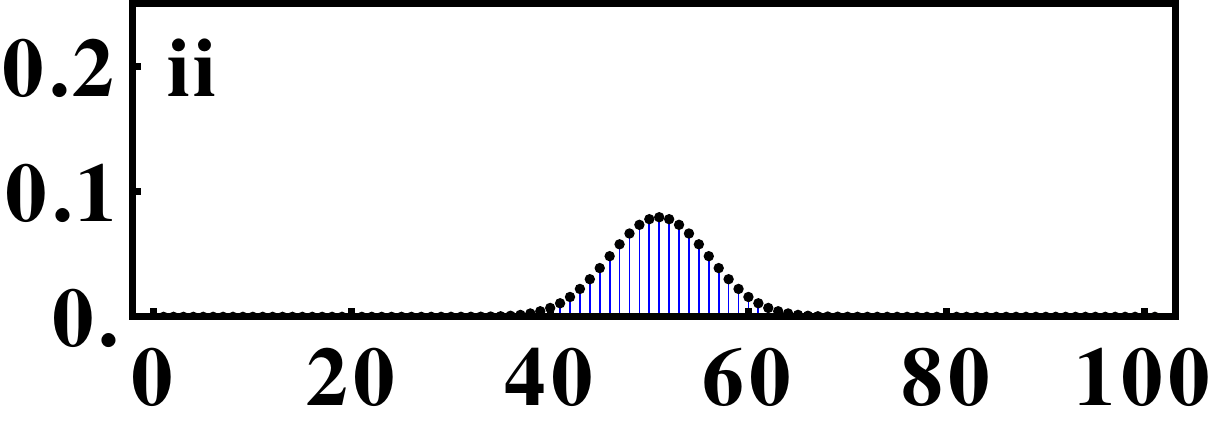}\enskip{}\,\quad{}\includegraphics[scale=0.4]{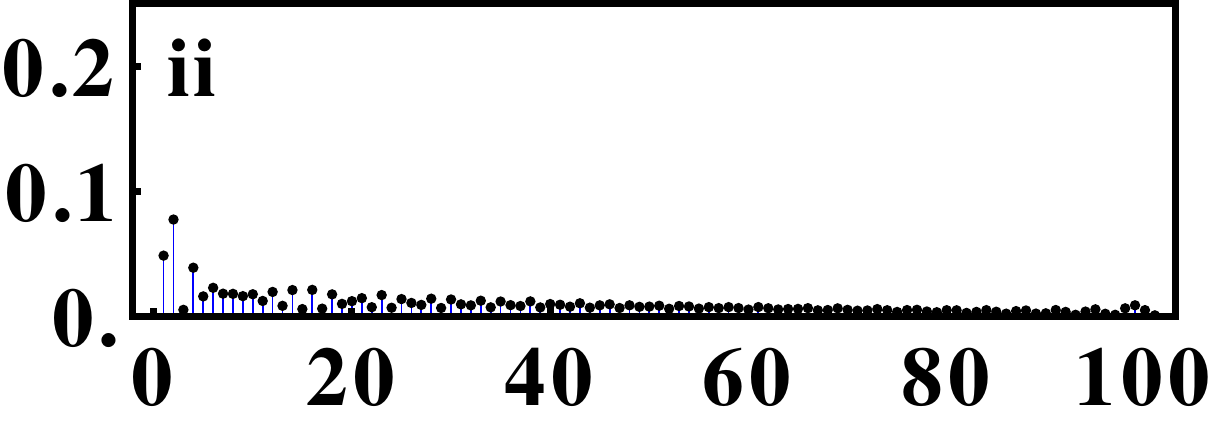} 
\par\end{centering}

\begin{centering}
\quad{}\enskip{}\includegraphics[scale=0.4]{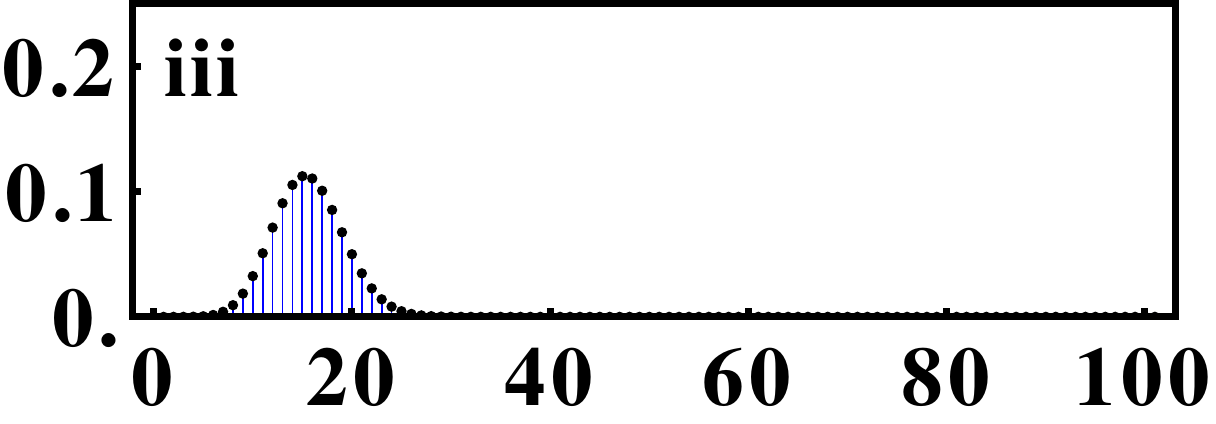}\enskip{}\,\quad{}\includegraphics[scale=0.4]{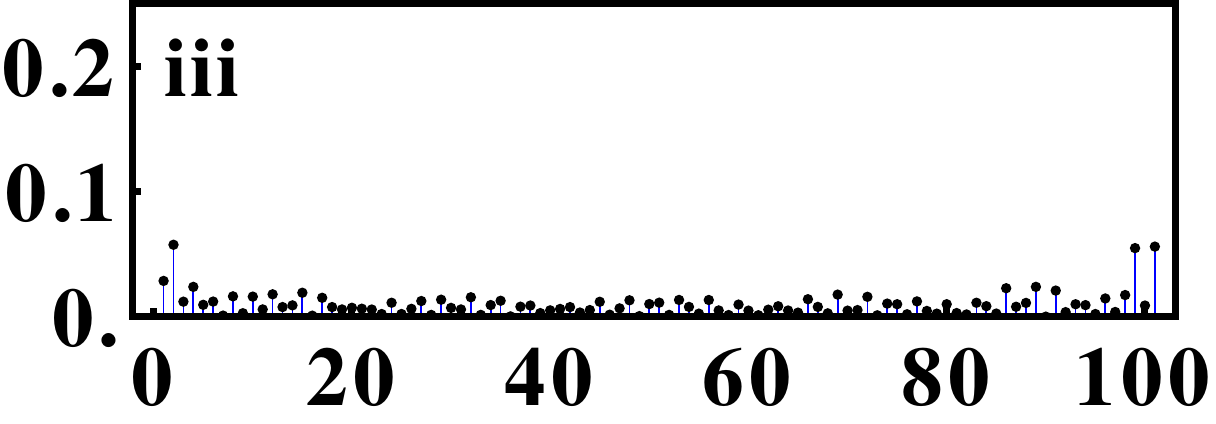} 
\par\end{centering}

\begin{centering}
\medskip{}
 \qquad{}(c) 
\par\end{centering}

\begin{centering}
\qquad{}\enskip{}$\left|C_{\phi}\right|^{2}$ vs $\phi/\pi$\enskip{}\quad{}\qquad{}\qquad{}\qquad{}\enskip{}$\left|C_{\phi}\right|^{2}$
vs $\phi/\pi$ 
\par\end{centering}

\begin{centering}
\quad{}\enskip{}\includegraphics[scale=0.4]{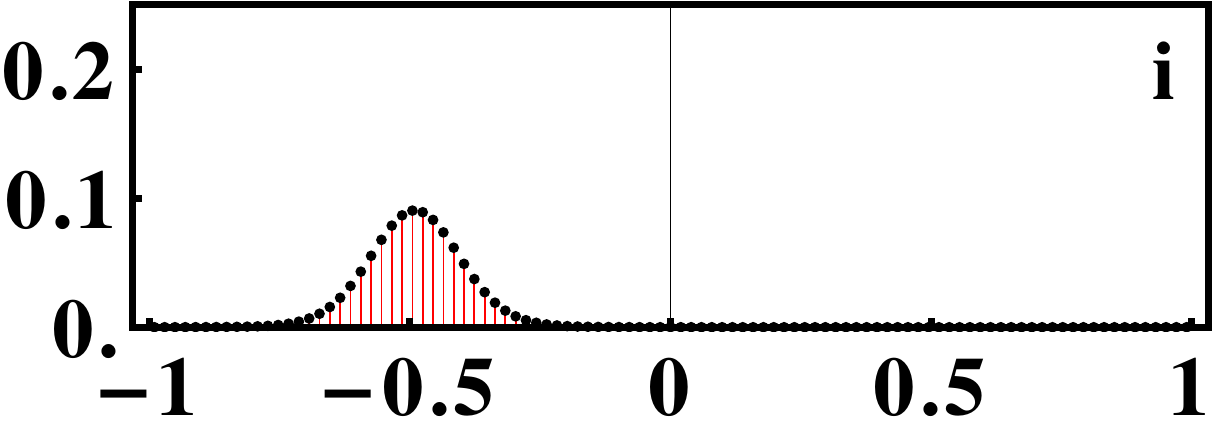}\enskip{}\,\quad{}\includegraphics[scale=0.4]{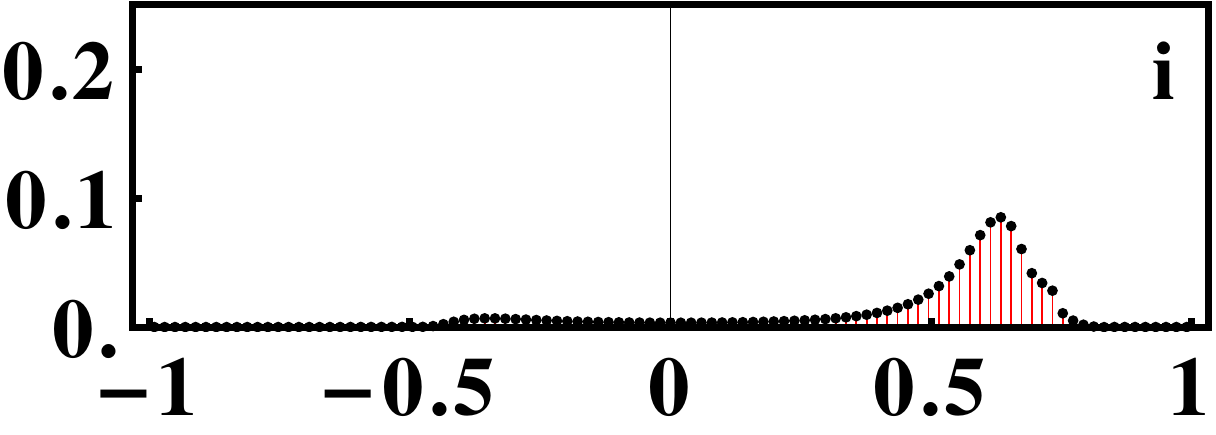}\\
 \quad{}\enskip{}\includegraphics[scale=0.4]{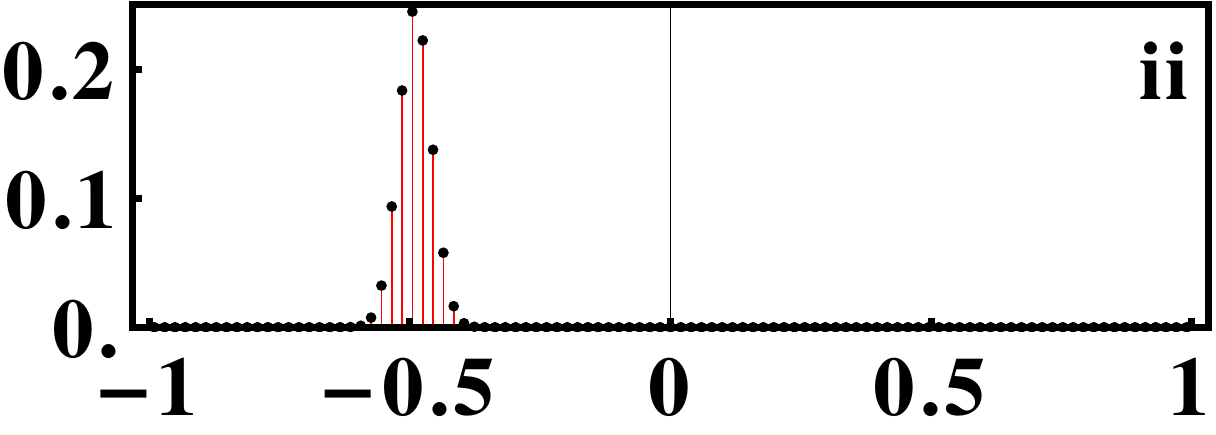}\enskip{}\,\quad{}\includegraphics[scale=0.4]{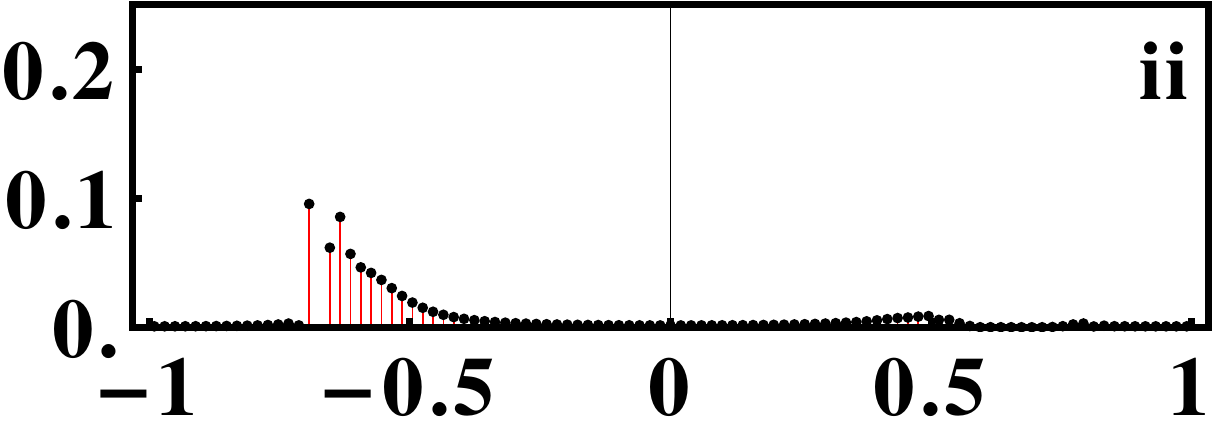}\\
 \quad{}\enskip{}\includegraphics[scale=0.4]{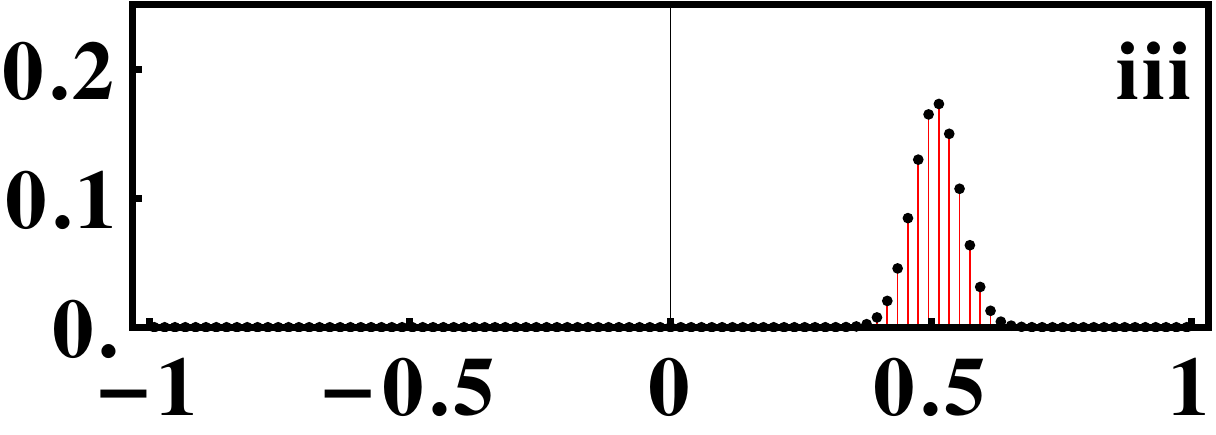}\enskip{}\,\quad{}\includegraphics[scale=0.4]{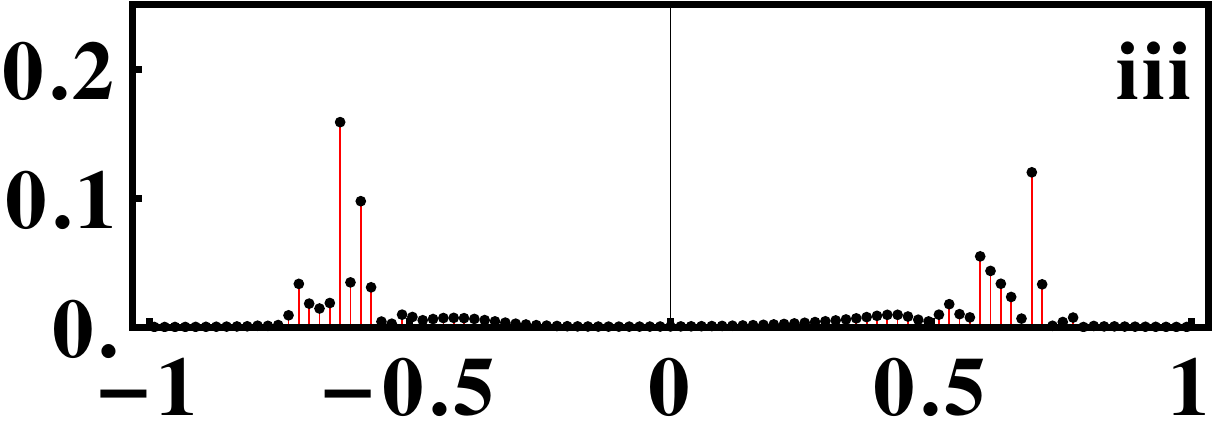}\\

\par\end{centering}

\caption{\label{fig:2}(color online) Comparison of the Rabi ($N\kappa_{0}=0$,
left column) and Josephson ($N\kappa_{0}=0.5\Omega$, right column)
dynamics, for an initial $\left|N\right\rangle $ Fock state preparation
(all $N=100$ bosons in one well) without external driving, illustrating
the process of collapse or {}``diffusion'' in relative number. (a)
The evolution of the mean relative number $\left\langle \hat{J}_{z}\right\rangle $
(blue thick line) and its standard deviation (black thin line); note
how in the Josephson case, the dispersion increases until it reaches
a saturation value close to that of a uniform distribution. For the
three sample times (i-iii), the relative number (b) and phase (c)
distributions are depicted.}
\end{figure}

To gain insight into the collapsing process, we start by considering
the non-interacting, non-driving case ($\kappa_{0}=0$, $\mu=0$:
Rabi regime) in (\ref{eq:2B-H_Hamiltonian}) , preparing a system
of $N=100$ atoms initially in the Fock state $\left|100\right\rangle $
which corresponds to all bosons located initially in one well, with
a $Q$-distribution centered around $\theta=0$, the north pole of
the Bloch sphere. As mentioned earlier this is also a coherent state,
the relative phase distribution is completely undetermined (uniform
distribution) and correspondingly its distribution in relative number
is sharp (delta distribution). In this situation, as is well known
\cite{RSF+99}, the boson population performs Rabi oscillations between
the wells at the tunneling frequency (Fig \ref{fig:2}(a), left).
Since the evolution operator is simply $\exp(+i\hat{J}_{x}\Omega t)$,
the result of its actuation on a coherent state is to produce another
coherent state rotated an angle $\Omega t$ around the $x$-direction
in the Bloch sphere, the relative number distribution, Fig. \ref{fig:2}(b),
changes accordingly going from the delta function at the poles to
a Gaussian shape at the equator, with its mean changing in time as
$N\cos^{2}(\theta(t)/2)$ and its variance as $N\sin^{2}(\theta(t)/2)\cos^{2}(\theta(t)/2)$,
with $\theta(t)=\pi-\Omega t$, as expected for a binomial distribution.
The phase distribution in Fig. \ref{fig:2}(c) evolves from uniform
to binomial distributions centered around $\pm\pi/2$ alternately,
as the coherent state goes from west to east of the sphere (since
the center of the Husimi distribution always lies on the $yz$-plane),
its width changing reciprocally with the width of the number distribution.

This periodic scenario changes when interactions are taken into account.
The evolution operator has now a complicated form since is composed
of the tunneling and nonlinear interaction operators, characterized
by $\hat{J}_{x}$ and $\hat{J}_{z}^{2}$ respectively, that do not
commute, and make it very difficult to evaluate its action analytically
in the Josephson regime. When the interactions are not so large compared
with the tunneling ($N\kappa_{0}/\Omega<1$), so the boson population
is not self-trapped, or equivalently the Husimi distribution is not
confined to the poles of the Bloch sphere, the dynamics is still dominated
by $\hat{J}_{x}$ but the different Fock states that compose the coherent
state will gain a phase $(n-N/2)^{2}$ due to the action of $\hat{J}_{z}^{2}$.
This dephasing will cause a spreading in the relative number distribution,
its mean tending to zero (a balanced population state) and its variance
increasing until it saturates to a value approximately equal to the
corresponding to a uniform distribution. This is what is called the
\textit{collapse} of the population oscillations (see the right column
of Figs. \ref{fig:2}(a) and (b)); it is a direct consequence of the
discreteness of the quantum state and is not captured by a mean-field
theory which predicts large amplitude anharmonic oscillations between
the wells \cite{RSF+99}. The phase distribution shows an interesting
behavior, since although for short evolution time it shows a well
defined distribution, when the relative distribution is smeared out,
for later times it tends to a distribution with phase averaged out
to zero due to a double-peaked profile as shown at \textasciitilde{}$\pm$0.6$\pi$
in Fig. \ref{fig:2}(c). We want to note that this process known as
{}``phase diffusion'' in the literature, may be a diffusion in relative
number (as in this case), relative phase, or even both, depending
on the initial state and model parameters, since $\phi$ and $n$
are complementary variables.

\begin{figure}
\begin{centering}
(i)\quad{}\quad{}\quad{}\quad{}\quad{}\quad{}\quad{}\quad{}(ii)\quad{}\quad{}\quad{}\quad{}\quad{}\quad{}\quad{}(iii) 
\par\end{centering}

\begin{centering}
\includegraphics[scale=0.32]{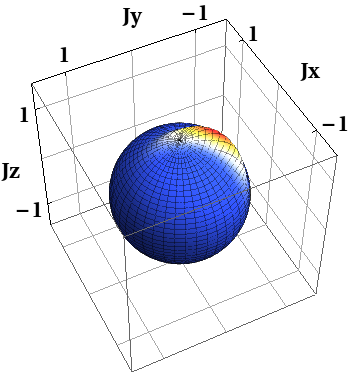}\includegraphics[scale=0.32]{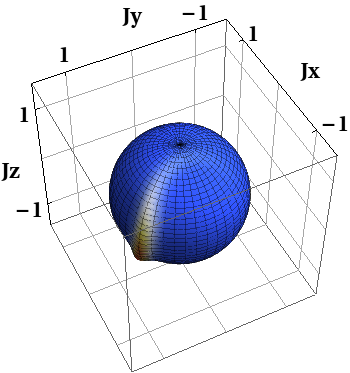}\includegraphics[scale=0.32]{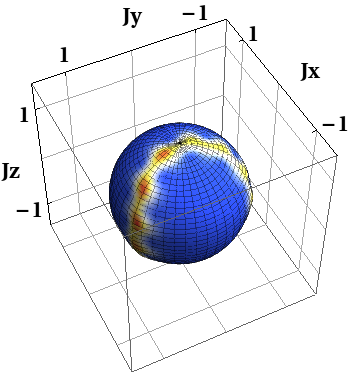} 
\par\end{centering}

\caption{\label{fig:3}(color online) Snapshots of the Husimi distribution
for three different instants (i-iii) as indicated in fig. \ref{fig:2}(a)
(right) with the same parameters and initial preparation, showing
the collapse of relative population on the phase space.}
\end{figure}

All those features can be better understood using the Fokker-Planck
equation obtained for the evolution of the Husimi distribution, $Q\equiv Q(\theta,\varphi;t)$
(see derivation in the Appendix), which behaves as a nonlinear fluid
moving on the spherical Bloch space: 
\begin{eqnarray}
\frac{\partial Q(\theta,\varphi;t)}{\partial t} & = & [-\Omega(\sin\varphi\frac{\partial}{\partial\theta}+\cot\theta\cos\varphi\frac{\partial}{\partial\varphi})\label{eq:FPEtext}\\
 &  & -2\kappa(N\cos\theta\frac{\partial}{\partial\varphi}-\sin\theta\frac{\partial^{2}}{\partial\theta\partial\varphi})]Q(\theta,\varphi;t).\nonumber 
\end{eqnarray}
 The tunneling term gives rise directly to the orbital angular momentum
operator $L_{x}$, thus having the effect of a drifting motion around
the $x$ direction, with a drifting coefficient proportional to the
tunneling frequency $\Omega$. This term only is the one responsible
for the dynamics in the Rabi regime, when no interactions are present
($\kappa=0$). On the other hand, interactions change the whole picture.
The interaction term is not translated simply into a $L_{z}^{2}$
operator, since it is produced by a nonlinear operator, yielding two
simultaneous effects on the dynamics. One is a drifting motion around
the $z$ direction produced by the first term of the second line of
Eq. (\ref{eq:FPEtext}), which can be seen is effected by a $L_{z}$
orbital angular momentum operator, with a $\cos\theta$-dependent
drifting coefficient, proportional to $N\kappa$; this indicates that
the effect of this drifting is stronger near the poles of the sphere
and negligible around the equator. The other is a diffusive effect,
produced by the cross-derivative in the last term of Eq. (\ref{eq:FPEtext}),
with a $\sin\theta$-dependent diffusion coefficient, proportional
to $\kappa$ - this term, when positive, spreads the distribution
hence being responsible for the \textit{loss of coherence}, but its
strength is much weaker than the drifting motion by a factor equal
to the number of particles, which can be very large. 
Notice that its action is out of phase with the interaction drifting,
so the diffusion is more acute in the equator, and also that it depends
in both angular variables, as was remarked earlier, for the case of
a general diffusion.

In Fig. \ref{fig:3} it is shown how the Husimi distribution evolves
under the Fokker-Planck equation (\ref{eq:FPEtext}) in the Josephson
regime ($N\kappa_{0}/\Omega<1$), for three sample times, as indicated
on the right column of Fig. \ref{fig:2}(a) %
\footnote{An animation with the complete dynamics shown in Fig. \ref{fig:3}
is available by request to the authors.%
}. Since for this set of parameters the tunneling is still dominating,
the distribution will rotate around the $x$-direction but now its
center will not stay on the $yz$-plane as in the Rabi case due, of
course, to the interactions. The latter will cause a twisting around
the $z$-axis (drift), and a spreading (diffusion) which generates
this strip-like shape. Because the motion depends on the polar coordinate
$\theta$, the spreading also occurs in this direction, which causes
the loss of definition in relative number. The two peaks of the phase
distribution on \textasciitilde{}$\pm$0.6$\pi$ in Fig. \ref{fig:2}(c)
are then understood since the Husimi distribution in this case is
well localized in the $\varphi$ direction around these values. If
the interaction term were the dominating one ($N\kappa\gg\Omega$:
Fock regime), the delocalization in phase space would be in the $\varphi$
direction instead because of the drifting motion around $z$, and
then the diffusion would be mostly in relative phase. In this regime,
depending on the initial conditions, the slow drifting in the periodic
surface around $x$ would cause the distribution to have enough time
to accumulate recovering coherence. The exact coordinates of this
coherence revival is completely defined by the interplay of the diffusion
and drifting mechanisms.

It is interesting to note how the classical transition point where
self-trapping starts to appear, is captured naturally from the Fokker-Planck
equation (\ref{eq:FPEtext}). Ignoring the diffusion term which is
negligible when $N\rightarrow\infty$ (since in this case the mean-field
and quantum approaches coincide), and comparing the two remaining
terms: the interaction dominates over the tunneling for values of
$N\kappa_{0}>\Omega/2$ (In the classical dynamics analysis this is the pitchfork
bifurcation point, \textit{cf.} eq. (47) from Ref. \cite{HS01} and following discussion).

\section{Dynamical scattering length: control of coherence}

\begin{figure}
\begin{centering}
\includegraphics[scale=0.45]{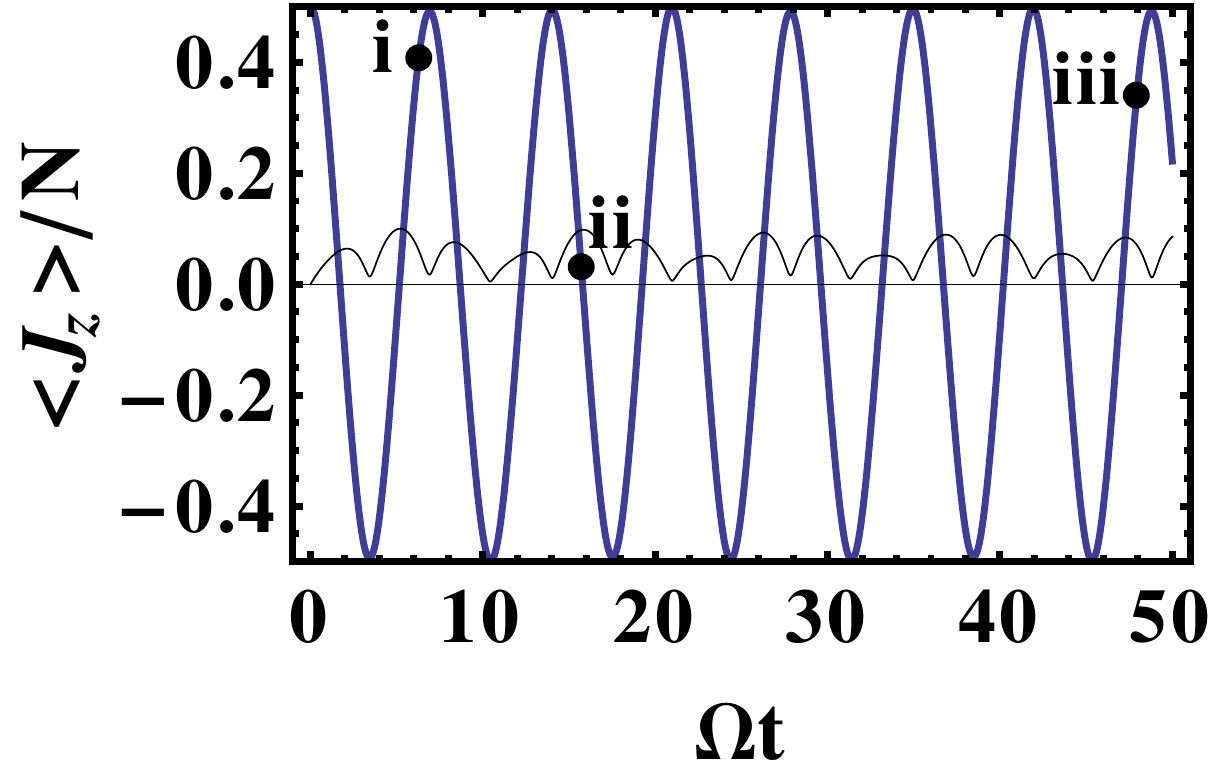} 
\par\end{centering}

\begin{centering}
\medskip{}

\par\end{centering}

\begin{centering}
(i)\quad{}\quad{}\quad{}\quad{}\quad{}\quad{}\quad{}\quad{}(ii)\quad{}\quad{}\quad{}\quad{}\quad{}\quad{}\quad{}(iii) 
\par\end{centering}

\begin{centering}
\includegraphics[scale=0.32]{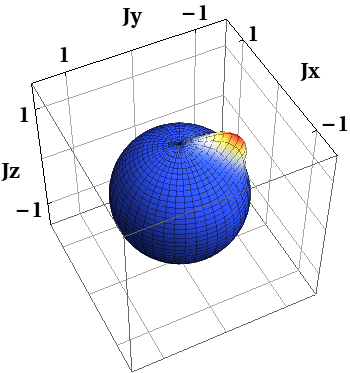}\includegraphics[scale=0.32]{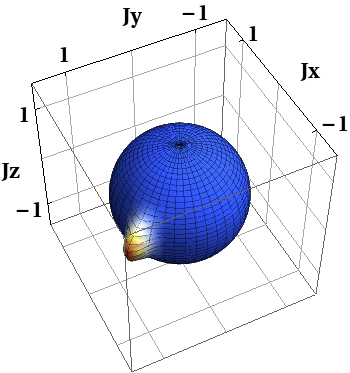}\includegraphics[scale=0.32]{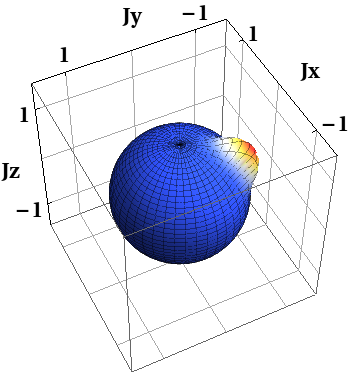} 
\par\end{centering}

\caption{\label{fig:4}(color online) Top: Dynamic control of coherence for
an initial atomic coherent state $\left|N\right\rangle $ with $N=100$
bosons, with interaction energy $N\kappa_{0}=0.5\Omega$, using the
driving parameters $\mu=0.3$ and $\omega=1.8\Omega$. The relative
population (thick blue) oscillates without collapsing in a very similar
fashion to the Rabi regime case. Bottom: The Husimi representation
for the state at three sample times (i-iii) as indicated in the above
figure, showing the localization in phase space of the distribution.}
\end{figure}

The preceding discussion sets the stage for the main result of this
paper, which is to control and utterly avoid the loss of coherence.
This occurs when we turn on the driving on the interactions, in a
certain region of the modulation parameters (amplitude and frequency),
for an initial atomic coherent state in the Josephson regime. In Fig.
\ref{fig:4} (top), for example, it is shown the evolution of the
mean relative number for the same initial state of Fig. 2, with a
driving frequency $\omega=1.8\Omega$ and a modulation amplitude $\mu=0.3$,
recalling that the interaction energy has the form $\kappa(t)=\kappa_{0}(1+\mu\cos\omega t)$,
as stated in Sec. II. It is remarkable the similarity of this case
with the non-interacting (Rabi) regime (c.f. the left column of Fig.
\ref{fig:2}(a), left), performing complete swapping of boson population
without collapsing to a balanced state, only that the period of the
oscillations is slightly larger. Also the standard deviation of the
relative number operator has the same behavior (black curve), reaching
small values when all the bosons are in one of the wells (i.e. the
poles of the Bloch sphere in the phase space representation) and its
largest values when the population is balanced (equator of the sphere),
but keeping always a much lower value than the corresponding to a
unifom distribution. Now, in the phase space picture (Fig. \ref{fig:4},
bottom), snapshots of the Husimi distribution for different times
in the evolution of the system show how the coherence is maintained,
since the distribution stays localized, deviating from a coherent
state just in that there is a squeezing in the relative phase. It
is possible to understand why the distribution keeps its coherence
directly from the Fokker-Planck equation. This occurs as the system
attains a resonant regime, which depends in a non-trivial manner
on $N$, $\Omega$, $\kappa_{0}$, as well as the strength $\mu$
and frequency $\omega$ of the scattering length modulation. In a
simplified ``averaged'' picture in this resonant regime the modulation contributes
by changing the diffusion coefficient in a commensurate manner with
the drifting around $x$, so that when the diffusion is higher the
drifting also is, so there is not ``enough time'' for the diffusion
to induce dephasing of the distribution components, which then keeps
coherence. There are fluctuations on this behavior since there is
diffusion and also drifting around the $z$-axis, which contribute
with this averaged picture to the overall dynamics, but they are small enough in a sense that the state is kept close to a coherent one. We performed numerical
tests for a longer time ($\Omega t$\textasciitilde{}500) than shown
in the figure and have not observed attenuation or balancing of the
relative population %
\footnote{The overall evolution depicted in Fig. \ref{fig:4} is better appreciated
in the animation, which is available by request to the authors.%
}.

\begin{figure}
\begin{centering}
\includegraphics[scale=0.6]{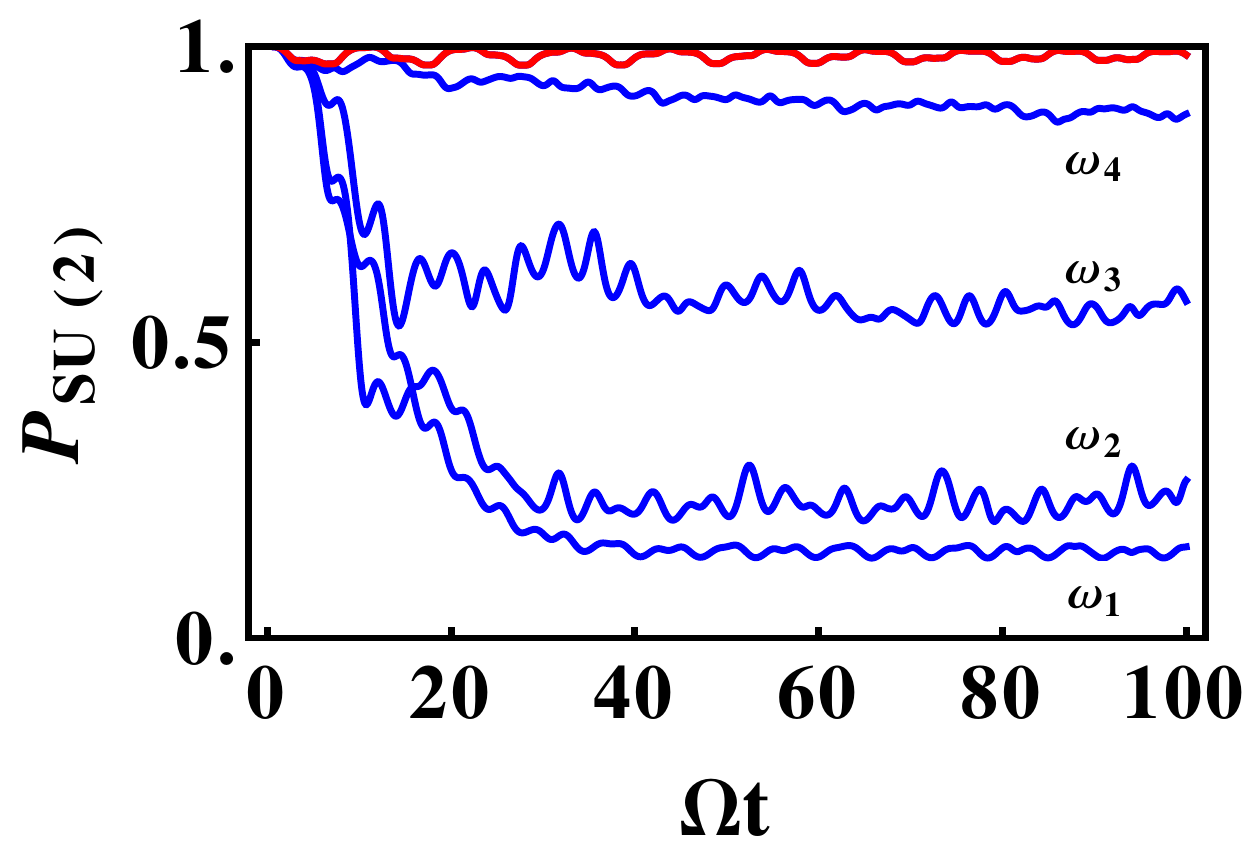}\quad{}\quad{}\quad{} 
\par\end{centering}

\caption{\label{fig:5}Temporal evolution of the generalized purity $\wp_{\,\,\textrm{SU(2)}}$
(or squared visibility $\mathcal{V}^2$) for the same initial preparation as
in fig. \ref{fig:4}, with driving frequencies: $\omega_{1}=1.0$,
$\omega_{2}=1.2$, $\omega_{3}=1.4$, $\omega_{4}=1.6$ (blue curves)
below the critical value $\omega_{\textrm{c}}\thickapprox1.8$ (red
curve), indicating the approaching of the transition where the collapse
of the semiclassical distribution is avoided.}
\end{figure}

For further confirmation of this maintenance of coherence, it was
used the generalized purity of the SU(2) algebra as a measure of the
separation of a given quantum state of our two-mode Hilbert space
from the coherent states inside that same space \cite{barnum,klyachko,viscondi},
defined by 
\begin{equation}
\wp_{\,\,\textrm{SU(2)}}(\left|\psi\right\rangle )=\frac{\sum_{i=x,y,z}\left\langle \psi|\hat{J}_{i}|\psi\right\rangle ^{2}}{N^{2}/4},\label{purity}
\end{equation}
 which attains its maximum value (of one) \textit{only} when $\left|\psi\right\rangle $
is an atomic coherent state, and decreasing when it starts to deviate
from it. From the phase space point of view, this deviation is greater
when the state gets more delocalized over the Bloch sphere, for example,
when the state of the system collapses in relative phase or number.
Interestingly, the generalized purity can be related to the fringe
visibility $\mathcal{V}$, of the interference pattern formed when
the fragmented condensate is released from the trap and let to expand
freely. The visibility is in fact given by $\mathcal{V}=\sqrt{\wp_{\,\,\textrm{SU(2)}}}$
(see the discussion concerning eq. (12) of \cite{G12} and compare
with our eq. (\ref{purity})). A high visibility is always desired
in the context of BEC interferometry since it translates into a better
resolution for phase measurements, and so is the generalized purity.
Maximal visibility (of one) is attained for an atomic coherent state.
However the same non-linearities due to interactions that make the
condensate useful for quantum interferometry (due to squeezing) \cite{GZN+10,GHM+10},
tend to diminish the visibility. This aspect is changed with the presence
of a modulation in the scattering length. In fig. \ref{fig:5}, the
evolution of $\wp_{\,\,\textrm{SU(2)}}$ is plotted for different
driving frequencies (blue) approaching from below the {}``resonance''
value (red) used in fig. \ref{fig:4}, with all the other parameters
and initial state fixed. It is clear that exists a critical
frequency which maintains the purity (and visibility) close to one
which for the initial Fock-coherent state $\left|N\right\rangle $
is around the value 1.8$\Omega$. Though not shown in the figure,
the purity also starts decreasing for higher frequencies.

\begin{figure}
\begin{centering}
\includegraphics[scale=0.5]{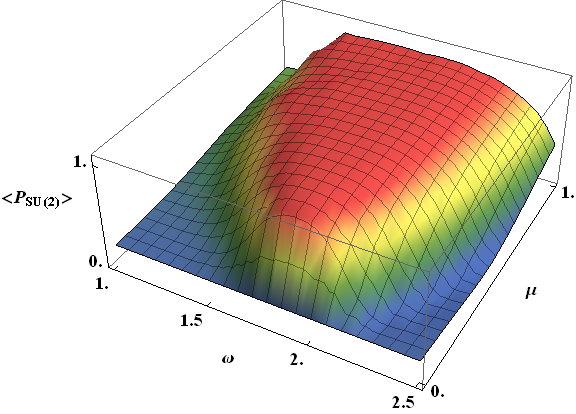} 
\par\end{centering}

\caption{\label{fig:6}(color online) Modulation parameters space diagram for
the temporal average of the SU(2) purity measure (or squared visibility $\mathcal{V}^2$), taken during a time
interval of $\Omega t=100$. The same initial state of fig. \ref{fig:4}
was used. Note the weak dependence on the modulation amplitude $\mu$.}
\end{figure}

In order to analyze the robustness of this control and the influence
of the amplitude of the modulation, it is plotted in Fig. \ref{fig:6}
the temporal average of the purity as a function of the driving parameters
$\omega$ and $\mu$, again for the same initial preparation; the
time average was taken for an interval of 100 in units of $\Omega^{-1}$.
It can be seen that effectively the maximum of the purity is reached
for the critical frequency ($\omega=$1.8$\Omega$ in this particular
case) and also the interesting feature that the control of coherence
is not strongly dependent on the variation of the amplitude $\mu$,
showing the same result for modulations going from 10 to 100\% of
the on-site static interaction $\kappa_{0}$. Other initial preparations,
less conventional from an experimental point of view were considered,
obtaining drastically different parameter diagrams indicating the
chaoticity of our model owing to the non-linear interaction term in
the Hamiltonian (\ref{eq:HamJ}) and moreover its explicit time dependence. Those features are responsible for a high unpredictability of the stability regime of parameters. 
Nonetheless, the procedure of using the purity or equivalently
the visibility has shown to be a valuable resource for finding regions
of stability and can be employed in any practical scheme, since it is associated to standard procedures in BEC interferometry experiments. 

\section{{\normalsize Conclusions}}

In conclusion, by studying a scattering length modulated periodically
in time it was found -for the case of a BEC in a double-well potential
inside the Josephson regime- a possible way to control the phase and
number diffusion of an initial coherent state. This avoidance of quantum
collapse was visualized with the assistance of the $Q$-Husimi distribution,
with the external driving field restraining the quantum distribution
from spreading over the Bloch sphere phase space. This was further
confirmed by the evolution of the generalized purity as a measure
of the distance from a coherent state, which is directly related to
the fringe visibility of the double-well interference. It was shown
that this kind of control is strongly dependent on the driving frequency,
with only a small influence from the amplitude of the modulation.
This feature of a driven Bosonic Josephson junction can be of great
interest for atom interferometry where various schemes for using this
two-mode systems have been suggested and implemented. The reason is
that by applying this technique, it is possible to turn off the undesired
effects caused by the nonlinear interactions, allowing for longer
holding times in the phase accumulation stage of the interferometer.
We want to note that though the initial preparation shown in this
paper is for the all-bosons-in-one-well Fock-coherent state, similar
regions of stability can be found for other preparations as for example
the typical balanced coherent state $\left|N/2\right\rangle $. Currently
interactions are the principal limitation towards actual Heisenberg-limited
metrology based on BECs. A future aspect to be considered is the extension of the present discussion for interferometry involving multiple traps, which has shown several promising features.

\ack J. L-V wishes to thank the Brazilian agency CNPq and the Emergent Leaders of the Americas Program (ELAP) through the Canadian Bureau of International Education (CBIE) for the financial support, and also the kind hospitality of Prof. David Feder at the Institute for Quantum Information Science at the University of Calgary where part of this work was done. MCO acknowledges support from AITF
	and the Brazilian agencies CNPq and FAPESP
	through the Instituto Nacional de Ci{\^e}ncia e Tecnologia em Informa\c{c}{\~a}o Qu{\^a}ntica (INCT-IQ).

\appendix

\section{Derivation of a Fokker-Planck equation for the Husimi $Q$-distribution
on the Bloch sphere}

\label{App}

We start from the von Neumann-Liouville equation 
\begin{equation}
\frac{\partial\hat{\rho}}{\partial t}=-i[\hat{H},\hat{\rho}],\label{eq:Liouville_eq}
\end{equation}
 with $\hat{H}$ given in (\ref{eq:2B-H_Hamiltonian}), and the goal
is to obtain an equation for the time-evolution of the Husimi $Q$-distribution
as a function of the angles $(\theta,\varphi)$ on the Bloch sphere.
First we take the mean of equation (\ref{eq:Liouville_eq}) on a coherent
state (\ref{eq:coherent_state}), 
\begin{eqnarray}
\frac{\partial Q}{\partial t} & = & -i\kappa\left\langle \theta,\varphi|[\hat{n}_{1}\left(\hat{n}_{1}-1\right)+\hat{n}_{2}\left(\hat{n}_{2}-1\right),\hat{\rho}]|\theta,\varphi\right\rangle \nonumber \\
 &  & +i\frac{\Omega}{2}\left\langle \theta,\varphi|[\hat{a}_{1}^{\dagger}\hat{a}_{2}+\hat{a}_{1}\hat{a}_{2}^{\dagger},\hat{\rho}]|\theta,\varphi\right\rangle .\label{eq:Qed}
\end{eqnarray}
 The first term of this equation, corresponding to the \textit{interaction},
can be rewritten as 
\begin{equation}
2i\kappa(N\left\langle \theta,\varphi|[\hat{n},\hat{\rho}]|\theta,\varphi\right\rangle -\left\langle \theta,\varphi|[\hat{n}^{2},\hat{\rho}]|\theta,\varphi\right\rangle )\label{eq:int_term}
\end{equation}
 with the definitions, $\hat{n}_{1}\equiv\hat{n}$ and $\hat{n}_{2}\equiv N-\hat{n}.$
Since in the Fock basis the density operator has the form $\hat{\rho}=\sum_{n,m=0}^{N}c_{n}c_{m}^{*}\left|n\right\rangle \left\langle m\right|,$
then the first commutator in (\ref{eq:int_term}) is evaluated as
\begin{equation}
[\hat{n},\hat{\rho}]=\sum_{n,m=0}^{N}c_{n}c_{m}^{*}(n-m)\left|n\right\rangle \left\langle m\right|,
\end{equation}
 and using the fact that 
\begin{eqnarray}
\left\langle \theta,\varphi\left|n\right\rangle \left\langle m\right|\theta,\varphi\right\rangle  & = & \sqrt{{{N}\choose{n}} {{N}\choose{m}}}e^{i(n-m)\phi}\cos^{n+m}(\frac{\theta}{2})\;\sin^{2N-(n+m)}(\frac{\theta}{2}),\label{eq:nm}
\end{eqnarray}
 it is readily obtained the following identity: 
\begin{equation}
\left\langle \theta,\varphi|[\hat{n},\hat{\rho}]|\theta,\varphi\right\rangle =-i\frac{\partial}{\partial\varphi}Q(\theta,\varphi;t).\label{eq:com}
\end{equation}
 A similar relation can be found for the anti-commutator, 
\begin{equation}
\{\hat{n},\hat{\rho}\}=\sum_{n,m=0}^{N}c_{n}c_{m}^{*}(n+m)\left|n\right\rangle \left\langle m\right|,
\end{equation}
 taking the derivative of (\ref{eq:nm}) with respect to $\theta$
and rearranging terms: 
\begin{eqnarray}
\left\langle \theta,\varphi|\{\hat{n},\hat{\rho}\}|\theta,\varphi\right\rangle  & = & [N(1+\cos\theta)-\sin\theta\frac{\partial}{\partial\theta}]Q(\theta,\varphi;t)\nonumber \\
\label{eq:anti}
\end{eqnarray}
 Using eqs. (\ref{eq:com}) and (\ref{eq:anti}) in the second commutator
in (\ref{eq:int_term}) 
\begin{equation}
[\hat{n}^{2},\hat{\rho}]=\sum_{n,m=0}^{N}c_{n}c_{m}^{*}(n-m)(n+m)\left|n\right\rangle \left\langle m\right|,
\end{equation}
 it is obtained the identity 
\begin{equation}
\left\langle \theta,\varphi|[\hat{n}^{2},\hat{\rho}]|\theta,\varphi\right\rangle =-i\frac{\partial}{\partial\varphi}[N(1+\cos\theta)-\sin\theta\frac{\partial}{\partial\theta}]Q(\theta,\varphi;t).
\end{equation}
 With these two identities the interaction term (\ref{eq:int_term})
can be written as 
\begin{equation}
2\kappa(\sin\theta\frac{\partial^{2}}{\partial\theta\partial\varphi}-N\cos\theta\frac{\partial}{\partial\varphi})Q(\theta,\varphi;t).\label{eq:FPE_int}
\end{equation}

For the \textit{tunneling} term in (\ref{eq:Qed}) we proceed in an
analogous manner, and after some algebra we obtain 
\begin{equation}
-\Omega(\sin\varphi\frac{\partial}{\partial\theta}+\cot\theta\cos\varphi\frac{\partial}{\partial\varphi})Q(\theta,\varphi;t).\label{eq:FPE_tunn}
\end{equation}
 Note that this differential operator is exactly the orbital angular
momentum operator around the $x$ axis. Collecting these two last
results in the von Neumann-Liouville equation for $Q(\theta,\varphi;t)$,
we finally obtain the Fokker-Planck equation 
\begin{eqnarray}
\frac{\partial Q(\theta,\varphi;t)}{\partial t} & = & [-\Omega(\sin\varphi\frac{\partial}{\partial\theta}+\cot\theta\cos\varphi\frac{\partial}{\partial\varphi})\\
 &  & -2\kappa(N\cos\theta\frac{\partial}{\partial\varphi}-\sin\theta\frac{\partial^{2}}{\partial\theta\partial\varphi})]Q(\theta,\varphi;t).\nonumber 
\end{eqnarray}

\end{document}